\documentstyle[psfig,draft]{mn}

\def\gs{\mathrel{\raise0.35ex\hbox{$\scriptstyle >$}\kern-0.6em
\lower0.40ex\hbox{{$\scriptstyle \sim$}}}}
\def\ls{\mathrel{\raise0.35ex\hbox{$\scriptstyle <$}\kern-0.6em
\lower0.40ex\hbox{{$\scriptstyle \sim$}}}}

\title[An Optical-Infrared Study of Galaxies in A\,2218]
      {A Photometric Study of the Ages and Metallicities of Early-type
      Galaxies in A\,2218}

\author[Smail et al.]
       {Ian Smail,$^{\! 1}$ Harald Kuntschner,$^{\! 1}$ 
	T.\ Kodama,$^{\! 1}$ G.\,P.\ Smith,$^{\! 1}$ C.\ Packham,$^{\! 2}$ 
	\and  A.\,S.\ Fruchter$^{3}$ \& R.\,N.\ Hook$^{4}$
        \vspace*{1mm}\\
        $^1$ Department of Physics, University of Durham, South Road,
             Durham DH1 3LE\\
	$^2$ Isaac Newton Group, Apartado 321, 38780 Santa Cruz de La
	     Palma, Tenerife, Canary Islands, Spain\\
	$^3$ Space Telescope Science Institute, 3700 San Martin Drive,
	     Baltimore, MD21210, USA\\
	$^4$ Space Telescope -- European Coordinating Facility,
	     European Southern Observatory, Karl-Schwarzschild-Str.\ 2,\\
	     ~~D-85748 Garching b.\ M\"unchen, Germany}

\date{\fbox{\sc Submitted: \today}}

\pagerange{000--000}

\begin{document}

\maketitle

\begin{abstract}
We present deep optical and near-infrared imaging of the rich cluster
A\,2218 at $z=0.17$.  Our optical imaging comes from new multicolour {\it
Hubble Space Telescope} {\it WFPC2} observations in the F450W ($B_{450}$),
F606W ($V_{606}$) and F814W ($I_{814}$) passbands.   These observations
are complemented by deep near-infrared, $K_s$-band, imaging from the new
INGRID imager on the 4.2-m William Herschel Telescope.  This combination
provides unique high-precision multicolour optical-infrared photometry
and morphological information for a large sample of galaxies in the core
of this rich cluster at a lookback time of $\sim$3\,Gyrs.  We analyse the
$(B_{450}-I_{814})$, $(V_{606}-I_{814})$ and $(I_{814}-K_s)$ colours of
galaxies spanning a range of a factor of 100 in $K$-band luminosity in
this region and compare these with grids of stellar population models.
We find that the locus of the colours of the stellar populations in
the luminous ($\gs 0.5L^\ast_K$) early-type galaxies, both ellipticals
and S0s, traces a sequence of varying metallicity at a single age.
At fainter luminosities ($\ls 0.1L^\ast_K$), this sequence is extended
to lower metallicities by the morphologically-classified ellipticals.
However, the faintest S0s exhibit very different behaviour, showing a
wide range in colours, including a large fraction (30\%) with relatively
blue colours which appear to have younger luminosity-weighted ages for
their stellar populations, 2--5\,Gyrs. We show that the proportion of
these young S0s in the cluster population is consistent with the observed
decrease in the S0 population seen in distant clusters when interpreted
within the framework of a two-step spectroscopic and morphological
transformation of accreted spiral field galaxies into cluster S0s.
\end{abstract}


\section{Introduction}

Studies using photometric, and more recently spectroscopic,
observations of luminous elliptical galaxies in distant clusters
have suggested that their luminosity and colour evolution is modest and
consistent with passive evolution of stellar populations which are
formed at high redshifts, $z>2$--4 (Aragon-Salamanca et al.\ 1993;
Ellis et al.\ 1997; van Dokkum et al.\ 1998; Kodama et al.\ 1998;
Kelson et al.\ 2000).  

The relatively modest evolution seen in the luminous elliptical
population in clusters contrasts with the claims of strong evolution in
the morphological mix in these environments, specifically the ratio of
S0 (lenticular) to elliptical galaxies.  Using {\it Hubble Space
Telescope} ({\it HST}\,) imaging of ten clusters at $z>0.3$--0.6,
Dressler et al.\ (1997) have uncovered a rapid increase in the ratio of
S0 to elliptical galaxies towards the present day (see also Fasano et
al.\ 2000).  Assuming that the elliptical population is static, they
interpret this as a strong rise in the S0 population, from 10--20\% at
$z\sim 0.5$ (6\,Gyrs ago) to the 60\% seen in present-day rich clusters.
This apparent evolution applies to the cluster population brighter than
about $0.1 L^\ast_V$.

By including information from spectroscopic observations of galaxies
in ten clusters, Poggianti et al.\ (1999) proposed that the build-up
of S0 galaxies identified by Dressler et al.\ (1997) resulted from the
morphological transformation  into lenticulars of accreted field spiral
galaxies.  They further suggested that the presence of a large population
of red, apparently passive, galaxies with late-type morphologies in
the clusters, along with  absence of blue S0 galaxies, indicated that
the morphological transformation occurs on a longer timescale than
the decline in the star formation rates of the accreted field galaxies
(see also Kodama \& Smail 2000).  Thus they proposed that the galaxies
first suffered a decline in their star formation (probably resulting from
stripping of their gas reservoirs as they encounter the dense intracluster
medium within the cluster potential) and only then had their morphological
appearance transformed (perhaps by a completely separate process, e.g.\
Moore, Lake \& Katz 1998).

Nevertheless, if the morphological transformation occurs on a timescale of
$\ls 2$--3\,Gyrs after the galaxy's entry into the cluster it may still be
possible to identify subtle signatures of this previous activity in the S0
population.  Direct comparisons of the estimated stellar population ages
for the S0 and elliptical populations can be particular powerful in this
respect.  When comparing the composite spectra of the S0 and elliptical
populations in three clusters at $z=0.3$, Jones, Smail \& Couch (2000)
found no significant differences in the inferred luminosity-weighted ages
of the stellar populations in the two classes at a median luminosity
of $0.6 L^\ast_V$ and concluded that this supported the two-step
transformation model discussed above.  A similar degree of homogeneity
between the cluster E and S0 galaxies was reported by Ellis et al.\
(1997) from their study of the restframe $(U-V)$ colours of early-type
galaxies in three clusters at $z\sim 0.5$, although again the strongest
constraints were on the more luminous galaxies, $\gg 0.2 L^\ast_V$.
However, a more detailed spectral-line analysis of individual lenticular
and ellipticals in Fornax (Kuntschner 2000) has suggested that, in this
cluster at least, the low luminosity lenticulars ($\ls 0.1 L^\ast_V$)
exhibit typically younger luminosity-weighted ages for their stellar
populations than the ellipticals (see also Jorgensen 1999).

Work on the relative ages of cluster ellipticals and lenticulars
continues using new combinations of spectral line indices to break the
degeneracy between age and metallicity (Worthey 1994; Vazdekis \&
Arimoto 1999; Kuntschner 2000).  At higher redshift such studies can be
observationally expensive, requiring relatively large amounts of 4- and
8-m time and efficient multi-object spectrographs (Kelson et al.\ 1997;
Jorgensen et al.\ 1999; Ziegler et al.\ 2000).  However, recent
developments in stellar population modelling (e.g.\ Vazdekis et
al.\ 1997; Kodama \& Arimoto 1997; Bruzual \& Charlot 2000) have
suggested that it may be possible to break the Age--Metallicity
degeneracy using observationally cheaper photometric analyses if it
includes both optical {\it and} near-infrared ($H$ or $K$) photometry
(an idea originally proposed by Aaronson 1978, see Peletier \&
Balcells 1996 for an early demonstration).  In this paper we explore
this possibility using new, high-quality observations of the galaxy
population in the core of the rich cluster A\,2218 ($z=0.17$) from {\it
HST}\, and the new INGRID near-infrared imager on the William Herschel
Telescope (WHT).  We derive precise optical and optical-infrared
colours and morphologies for a large sample of galaxies across a 6
magnitude range in near-infrared luminosity down to $K_s=19$ ($0.02
L^\ast_K$) in this field and compare these to the predictions of the
latest stellar evolutionary models.  We also compare the constraints we
derive for individual galaxies with the recent spectroscopic study of
A\,2218 by Ziegler et al.\ (2000).  We discuss our observations and
their reduction in the next section, present our results in \S3 and our
conclusions in \S4.  Throughout we assume $q_o=0.5$ and $h=0.5$ in
units of 100\,km\,s$^{-1}$\,Mpc$^{-1}$.  In this cosmology the
look-back time to A\,2218 is 3\,Gyrs, the angular scale at the cluster
is 1$'' \equiv 4.3$\,kpc and $K^\ast_s=14.8$ (adopting $M^\ast_K=-25.7$).

%
%
\begin{figure*}
\vspace*{3.0in}
\centerline{\Huge Fig1.jpg}
\vspace*{3.0in}
~\bigskip
\caption{A true colour {\it HST}\, image of the core of A\,2218 composed
from the F450W (blue), F606W (green) and F814W (red) images.  North is
up and east is to the left.  We identify the galaxies employed in our
study using the numbering scheme given in Table~1.  The coordinates are
measured relative to the position of the central cD galaxy (\#301,
$\alpha=16^h\,35^m\,49^s\!\!.31$, $\delta=+66^\circ\,12'\,44"\!\!.7 $
(J2000)).}
\end{figure*}

\section{Observations, Reduction and Analysis}

\subsection{{\it HST}\, Optical Imaging}

The {\it HST}\footnote{This paper is based upon observations obtained
with the NASA/ESA {\it Hubble Space Telescope} which is operated by
STScI for the Association of Universities for Research in Astronomy,
Inc., under NASA contract NAS5-26555.}\, imaging analysed here was
obtained as part of the Early Release Observations (ERO) from {\it
WFPC2} after the SM-3a servicing mission in January 2000.   The
observations comprise 5 orbits in each of the F450W, F606W and F814W
passbands, giving total integration times of 11.0\,ks, 10.0\,ks and
12.0\,ks respectively.  These exposures were each broken into 1000\,s
sub-exposures which were spatially offset from each other by sub-pixel
shifts to allow drizzle-reconstruction of the frames to recover some of
the information lost due to the undersampling of the {\it HST}\, PSF by
the {\it WFC} pixel arrays.

The individual images were processed using the standard {\it WFPC2}
data pipeline and subsequently combined and cleaned of cosmic ray and
other artifacts using {\sc drizzle} and related software (Fruchter \&
Hook 1997) as implemented in {\sc iraf}/{\sc stsdas} .  The final {\it
WFC} images have an effective resolution (FWHM) of 0.17$''$,
0.05$''$-sampling and cover a field of 5.0 sq.\ arcmin.  We adopt the
photometric calibration of Holtzman et al.\ (1995) as given on the
STScI web pages to calculate magnitudes in the Vega-based
$B_{450}$/$V_{606}$/$I_{814}$ system.  We use an average Gain Ratio of
1.988 for the three {\it WFC} chips, but note that for the colours
based on the {\it HST}\, photometry the adopted Gain Ratio cancels out,
and for the {\it HST}--INGRID colours the uncertainty it introduces is
much smaller than the calibration error of the ground-based
photometry.  The 3-$\sigma$ point source sensitivities in the two
passbands are then:  $B_{450}=28.8$, $V_{606}=29.0$ and
$I_{814}=28.1$.  At the redshift of A\,2218 these filters sample the
cluster population in passbands roughly equivalent to restframe $U$,
$V$ and $R$.  We show a true-colour representation of the cluster as
seen in the {\it HST}\, imaging in Figure~1.

\subsection{INGRID Near-infrared Imaging}

Our near-infrared observations come from the new INGRID imager on the
4.2-m WHT\footnote{Based on observations made
with the William Herschel Telescope operated on the island of La Palma
by the Isaac Newton Group in the Spanish Observatorio del Roque de los
Muchachos of the Instituto de Astrofisica de Canarias} (Packham et
al.\ 2000).  These observations were obtained in commissioning time
during March 22--23, 2000.  INGRID comprises a 1024$^2$ HAWAII-2 array at
the bent-Cassegrain focus of the WHT, providing a 4.13\,arcmin field of
view with 0.242$''$/pixel sampling.  

Our exposures consist of a total of 8.3\,ks integration in the $K_s$
filter obtained under photometric conditions in $\sim
0.6$--0.8$''$ seeing.  An additional 9.1\,ks of $J$-band imaging
was also obtained and we discuss this elsewhere (Smith et al.\ 2000).
The $K_s$ frames consist of individual 20\,s exposures which are co-added
in hardware.  These were reduced in a standard manner using a smoothed
illumination correction derived from all the $K_s$ frames obtained during
the night and then a local sky correction constructed from a running
median of frames around a particular science frame.  The frames were
then aligned using integer-pixel shifts and combined using a cosmic-ray
rejection algorithm to produce a final frame.  The photometric calibration
of these frames is obtained from exposures of UKIRT Faint Standards
interspersed between the science observations.  We estimate that our
absolute calibration is good to 0.03\,mags (including the uncertainty
in the transformation from $K$ to $K_s$ for the faint 
standards, Persson et al.\ 1998).  The final stacked $K_s$-band
frame (approximately restframe $H$-band) has a 3-$\sigma$ point-source
sensitivity, within the seeing disk, of $K_s=22.0$ and seeing of 0.75$''$.

\subsection{Galaxy Catalogue and Colours}

To undertake our analysis we construct a $K_s$-selected galaxy
catalogue.  This achieves two goals -- it minimises our sensitivity to
differences in the previous star formation histories of the galaxies we
are studying, and as the $K_s$ frames have effectively the shallowest
limit, selecting in the $K_s$-band  guarantees a sample with the most
uniform optical-infrared colours and errors.  Looking at the model
grids shown in Figure~3 we can see that to differentiate between
different stellar populations we require typical photometric errors of
better than $\leq 0.1$\,mag in $(I_{814}-K_s)$.  This imposes an
effective $K_s$-band magnitude limit of $K_s<19.0$ on our sample,
equivalent to $M_{K^\ast}+4.2$ mags or roughly $M_K=-21.5$ for early-type
galaxies in A\,2218.

We therefore ran SExtractor (Bertin \& Arnouts 1996) on the final $K_s$
frame using a detection criteria of 10 pixels above the
$\mu_{K_s}=21.2$\,mag\,arcsec$^{-2}$ isophote (1.5$\sigma$).  To
determine the total magnitudes of galaxies on this frame we use the
{\sc best\_mag} estimate from SExtractor and confirm that this provides
a reliable ($\delta K_s \leq 0.05$) estimate of the total magnitudes
using large-aperture photometry of several bright, isolated galaxies in
the field.  The SExtractor  catalogue provides a list of 106 sources
within the joint {\it WFPC2}--INGRID field down to our adopted limit of
$K_s=19$.  We discard 10 sources as being stellar on the basis of their
morphologies in the F814W frame, a further 8 due to crowding (another
$K_s<19$ source within 1--2 half-light radii) and 7 which exhibit
gravitationally-lensed morphologies which obviously discount them from
our analysis of the cluster population (an optical-infrared photometric
analysis of the arcs and arclets seen in A\,2218 will be presented by
Smith et al.\ 2000).  This leaves us with a final sample of 81
candidate cluster galaxies which are brighter than $K_s=19$ and
relatively uncrowded, these galaxies are identified on Figure~1.

Morphologies of the galaxies in the A\,2218 field were kindly provided
by Prof.\ Warrick Couch based on the existing {\it WFPC2} imaging in
the F702W ($R_{702}$) passband presented by Kneib et al.\ (1996).
These morphologies are on the revised Hubble scheme as used by the
MORPHS project (Smail et al.\ 1997).  We have chosen to use visual
classifications for our analysis, rather than profile fitting, as all
of the claims for morphological evolution of the galaxy populations
of distant clusters have used visual estimates of the morphologies.
Therefore any signatures of this evolution should appear in visually
classified samples.  The roll-angle and field centre for the earlier
F702W observations were different from the ERO observations analysed
here and so there is a slight mismatch in the field coverage between the
two datasets.  This results in 15 galaxies out of the 81 galaxies in our
final $K_s<19$ sample lacking morphologies.  Morphological classification
of these galaxies was undertaken by one of us (IRS) from the F814W frame.
These galaxies are identified by ID's of 4000 and above in Table~1.
The morphologies of the $K_s<19$ sample break down as:  E: 31; E/S0:
3; S0:  28; Sa: 5; late-type spirals: 14.  The core of A\,2218 thus
exhibits an observed S0-to-E ratio, $N_{\rm S0}/N_{\rm E} \sim 1$, close
to that expected from the evolutionary rate proposed by Dressler et al.\
(1997), and a large deficit of S0 galaxies compared to the ratio seen in
local clusters, $N_{\rm S0}/N_{\rm E} \sim 2.3$.  Within the framework
of the S0-transformation model we would expect that roughly half of the
S0 population we see in A\,2218 had their morphologies transformed since
$z\sim 0.5$.

To determine representative colours for these galaxies we have chosen
to measure photometry from the seeing-matched frames within an aperture
whose radius is three times the half-light radius of the galaxy.   The
half-light radii, $r_{hl}$, we quote are simply the radius of a circle
containing half the total light of a galaxy on the F814W frame. To
estimate the total magnitudes we use the SExtractor {\sc best\_mag}
measurements and correct these to `total' magnitudes using
8.0$''$-diameter aperture photometry of isolated early-type galaxies on
the frame.  This correction amounts to $-0.05\pm 0.01$\,mag.  A
comparison of our half-light radii with the effective radii published
by Jorgensen et al.\ (1999) and Ziegler et al.\ (2000) from their
analyses of the {\it HST}\, F702W-band frame shows reasonable agreement,
$\ls 20$\% scatter.  The median half-light radius reaches 0.4$''$
(equivalent to the ground-based seeing) by our limit of $K_s=19$.  For
the galaxies with half-light radii below 0.4$''$ (14/81) we have chosen
to use a fixed aperture of 2.4$''$ diameter.

%
%
\centerline{\psfig{file=fig2.ps,angle=0,width=3.0in} }
\noindent{{\bf Figure 2.} Colour-magnitude diagrams for the $K_{s}<19$
sample showing their $(B_{450}-I_{814})$, $(V_{606}-I_{814})$ and
$(I_{814}-K_{s})$ colours as a function of total $K_{s}$ magnitude.
The points are coded on the basis of their optical morphology given
in the key.  Note the tight sequences exhibited by the colours of
the early-type galaxies, especially in $(V_{606}-I_{814})$.}

To measure the colours of the galaxies we aligned and resampled the
{\it HST}\, frames to the coordinate system defined by the INGRID $K_s$
frame to a tolerance of $\ll 0.1''$.  We then matched the effective PSF
on the {\it HST}\, and ground-based frames using an iterative
gaussian-convolution technique based on the PSF profiles within the
same $3 r_{hl}$ radius used in our analysis.  For each galaxy we
measure the $(B_{450}-I_{814})$, $(V_{606}-I_{814})$ and
$(I_{814}-K_{s})$ colours within the adopted apertures.  To judge the
sensitivity of our colours to the PSF correction we have reanalysed our
frames using PSF convolutions at either extreme of the range of
acceptable values and have determined that the $(I_{814}-K_s)$ colours,
which have the strongest sensitivity to this correction, vary by less
than 0.02\,mags.  We have also remeasured the colours using apertures
which are twice the half-light radii and determine that the
distribution of the early-type population on the colour--colour planes
does not significantly alter, with systematic changes in the
colours of $< 0.02$ mags.

We list the aperture photometry for our sample, along with the errors
estimated from photon-counting statistics and the variance of the sky
background, our best-estimate of the total $K_s$ magnitudes and the
half-light radii in Table~1. Table~1 also gives the positions of all
the galaxies, this astrometry is tied to the APM coordinate system
and has positional accuracy of $\ls 0.5''$ rms.  We illustrate the
colour-magnitude relations for the galaxies in the different passbands in
Figure~2. The estimated galactic extinction for this field from Schlegel
et al.\ (1998) is $E(B-V)=0.024$, giving $E(B_{450}-I_{814})=0.058$,
$E(V_{606}-I_{814})=0.034$, $E(I_{814}-K_{s})=0.038$ and $A_{K}=0.01$.
We correct all colours for extinction but have not corrected the total
$K_s$-band magnitudes.  Adopting the Burstein \& Heiles (1982) estimate
of the extinction reduces the reddening corrections by 25\% -- or at
most 0.01 mags.

\subsection{Archival Spectroscopy}

We expect that field galaxies are likely to comprise a relatively
modest fraction of our $K_s$-selected sample given the richness and
redshift of the cluster we are observing (especially for the early-type
galaxies).   Nevertheless, we have searched for spectroscopic
confirmation of the membership of the galaxies in our sample.  The
primary sources for this information are Le Borgne, Pello \& Sanahuja
(1992), Jorgensen et al.\ (1999) and Ziegler et al.\ (2000).  We find a
total of 33 of the 81 galaxies in our sample have spectroscopic
redshifts.   We list the identifications for these galaxies and their
redshifts in Table~1, following the naming convention from
NED\footnote{The NASA/IPAC Extragalactic Database (NED) is operated by
the Jet Propulsion Laboratory, California Institute of Technology,
under contract with the National Aeronautics and Space
Administration.}.

Of the 33 spectroscopically identified galaxies only two are
non-members, and only one of these (\#4010) has an early-type
morphology -- indicating a field contamination of 3\% (for the
typically brighter galaxies above the spectroscopic limits).   The
observed colours of \#4010 project it onto the sequence of metal-rich,
luminous ellipticals in the cluster.  Three other much fainter
ellipticals also appear in this region (identified by their large
photometric errors in Fig.~4c) and we suggest that these probably
represent background field galaxies, although this requires
spectroscopic confirmation.  The morphological breakdown of the
spectroscopic members is: 17/31 E's, 1/3 E/S0's, 13/28 S0's and 2/5
Sa's, showing that we have confirmed membership for roughly 40\% of the
early-type galaxies in our field.  In addition, 14 of the early-type
galaxies in our sample have high signal-to-noise, moderate resolution
spectroscopy available from Ziegler et al.\ (2000).  We discuss these
galaxies in more detail in the next section.

%
%
\setcounter{figure}{2}
\begin{figure*}
\centerline{\psfig{file=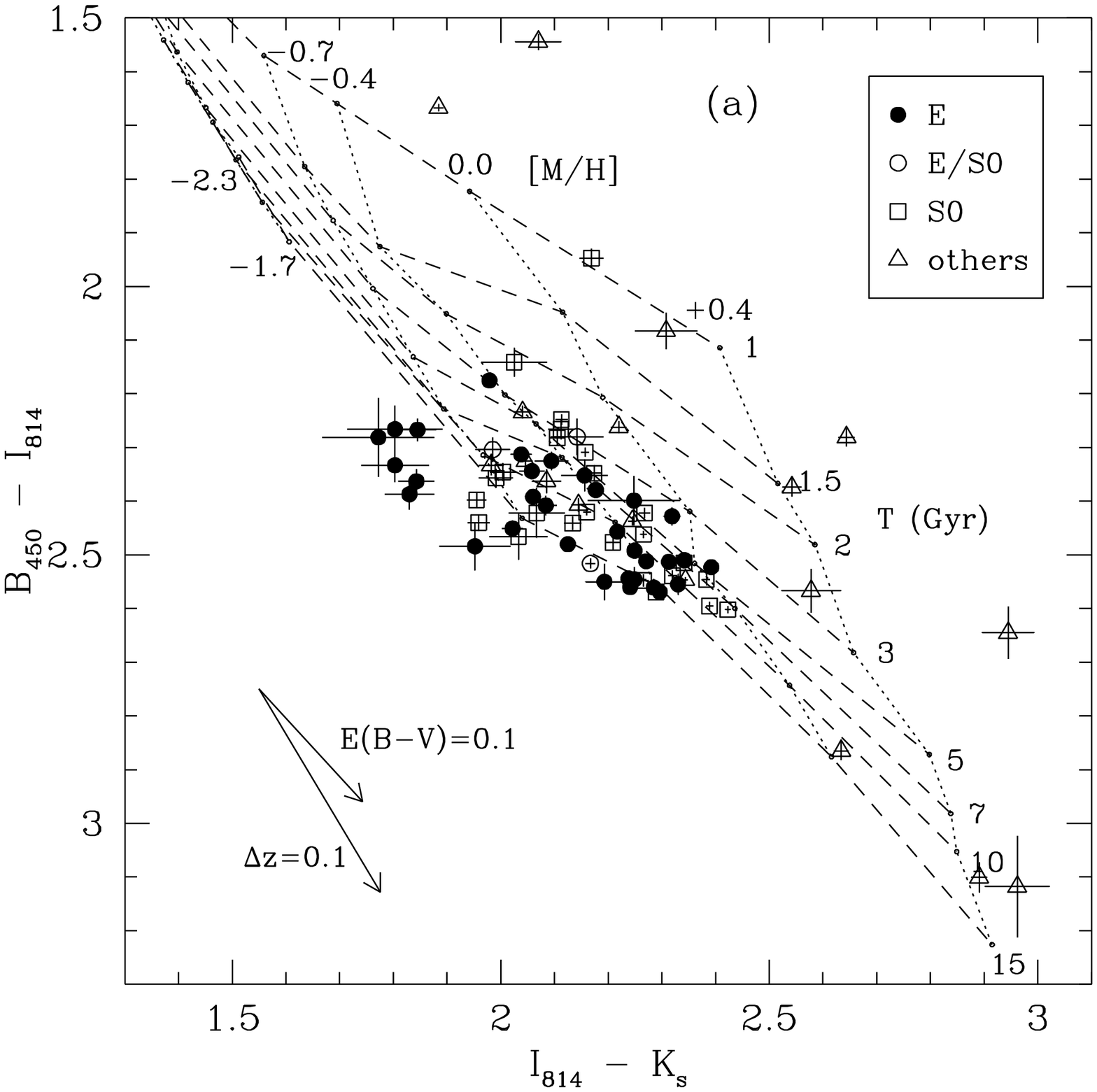,angle=0,width=3.0in} \hspace*{0.3in}
\psfig{file=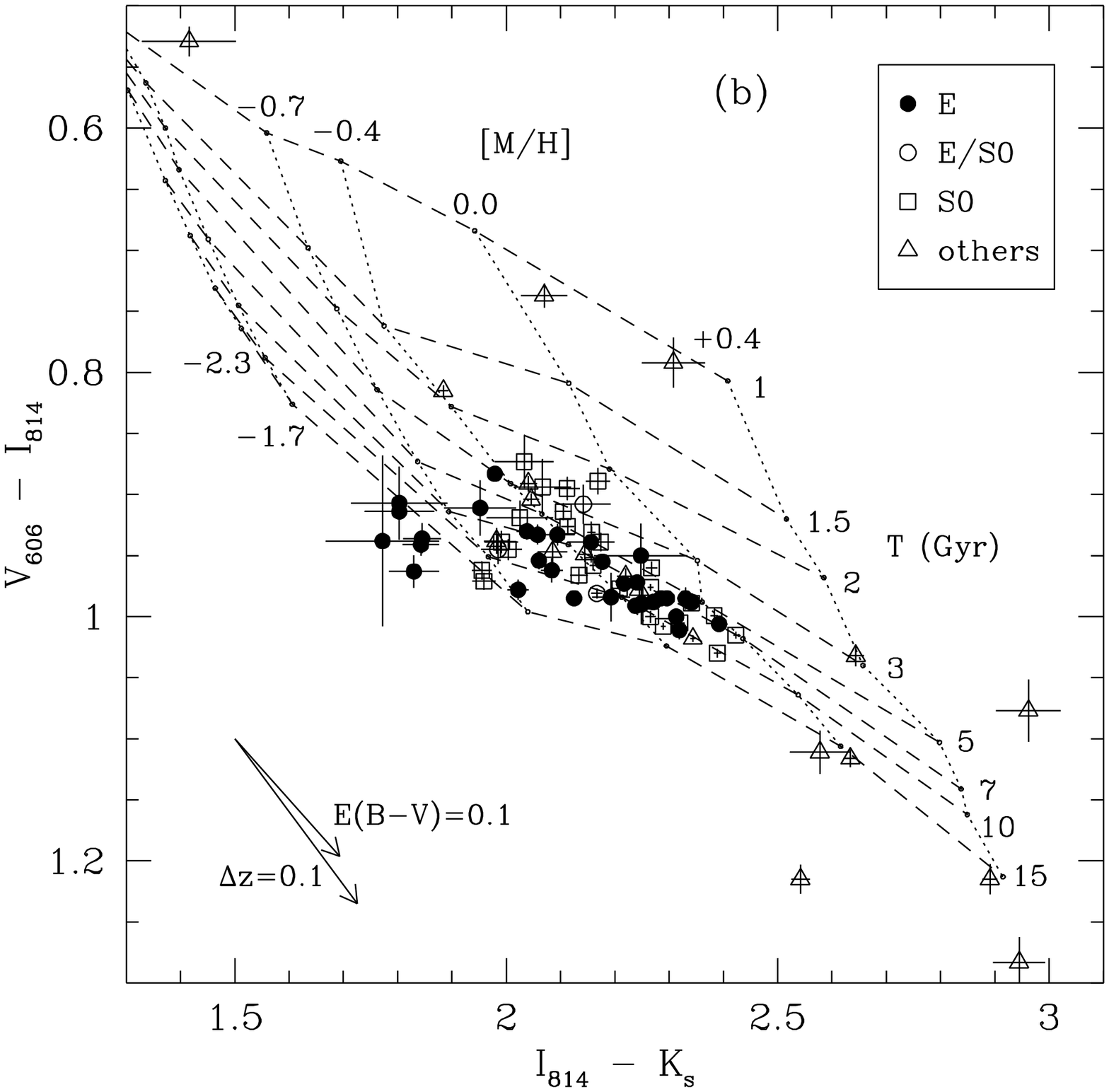,angle=0,width=3.0in}}
\caption{Colour-colour diagrams for the $K_{s}<19$ sample comparing the
$(B_{450}-I_{814})$ and $(V_{606}-I_{814})$ versus $(I_{814}-K_{s})$
planes.  The points are coded on the basis of their optical morphology
as given in the key.  The two vectors show the effects on the galaxy
colours of increasing the internal reddening and increasing redshift
(assuming an early-type spectral energy distribution).}
\end{figure*}

\subsection{Stellar Population Models}

We compare the observed colours of the galaxies with the predictions from
Bruzual \& Charlot's (2000) population synthesis code.  The single stellar
population (SSP, a stellar population with a single metallicity and age)
model grids are calculated for six different metallicities ([M/H]$= -$2.3,
$-$1.7, $-$0.7, $-$0.4, 0.0, and $+$0.4) and eight ages ($T=1$, 1.5, 2,
3, 5, 7, 10, 15 Gyrs).  We assume a Salpeter (1955) initial mass function
($x=1.35$) with lower and upper mass cut-offs of 0.1 and 100\,M$_{\odot}$.
The colours are calculated in the observer's frame at the cluster redshift
($z=0.17$) by redshifting the SSP spectra and convolving these with the
filter response functions.

As shown in Figure~3, the Age--Metallicity degeneracy can be broken by
combining a rest-frame optical colour and a rest-frame optical-infrared
colour. This is because the optical-infrared colour (e.g.\
$(I_{814}-K_s)$) of an old stellar population ($>1$\,Gyr) is primarily
determined by the temperature of the red giant branch which is much
more sensitive to the metallicity than the age of the stellar
populations (Table~2).  In contrast, an optical colour
(e.g.\ $(V_{606}-I_{814})$) is sensitive to {\it both} age and
metallicity in accordance with the predictions of Worthey's (1994) 2/3
law.  Bluer optical colours which sample the restframe near-ultraviolet
(e.g.\ $(B_{450}-I_{814})$ at $z=0.17$) are slightly more sensitive to
the metallicity than the 2/3 law suggests, since this colour brackets
the rest-frame 4000\AA\ break which is sensitive to metal lines
(e.g.\ Fe and Ca HK).  Therefore the best combination to break the
Age--Metallicity degeneracy for a cluster at $z\sim 0.2$ is
$(V_{606}-I_{814})$ versus $(I_{814}-K_s)$.

\setcounter{table}{1}
\begin{table}
\caption{The sensitivity to age and metallicity of the colours used in
our analysis. The rate of change of colour is determined for
$2<T ({\rm Gyr})<10$ and $-0.7<[{\rm M/H}]<0.0$.  The ratio,
($d$(col)/$d$[M/H])/($d$(col)/$d\log T$), is also shown in the final column.}
\begin{tabular}{lccc}
\hline
Colour & $d$(col)/$d$[M/H] & $d$(col)/$d\log T$ & Ratio \cr
\hline
$(B_{450}-I_{814})$  & 0.53 & 0.56 & 0.96 \cr
$(V_{606}-I_{814})$  & 0.15 & 0.22 & 0.67 \cr
$(I_{814}-K_s)$      & 0.76 & 0.45 & 1.73 \cr
\hline
\end{tabular}
\end{table}

%
%
\setcounter{figure}{3}
\begin{figure*}
\centerline{\psfig{file=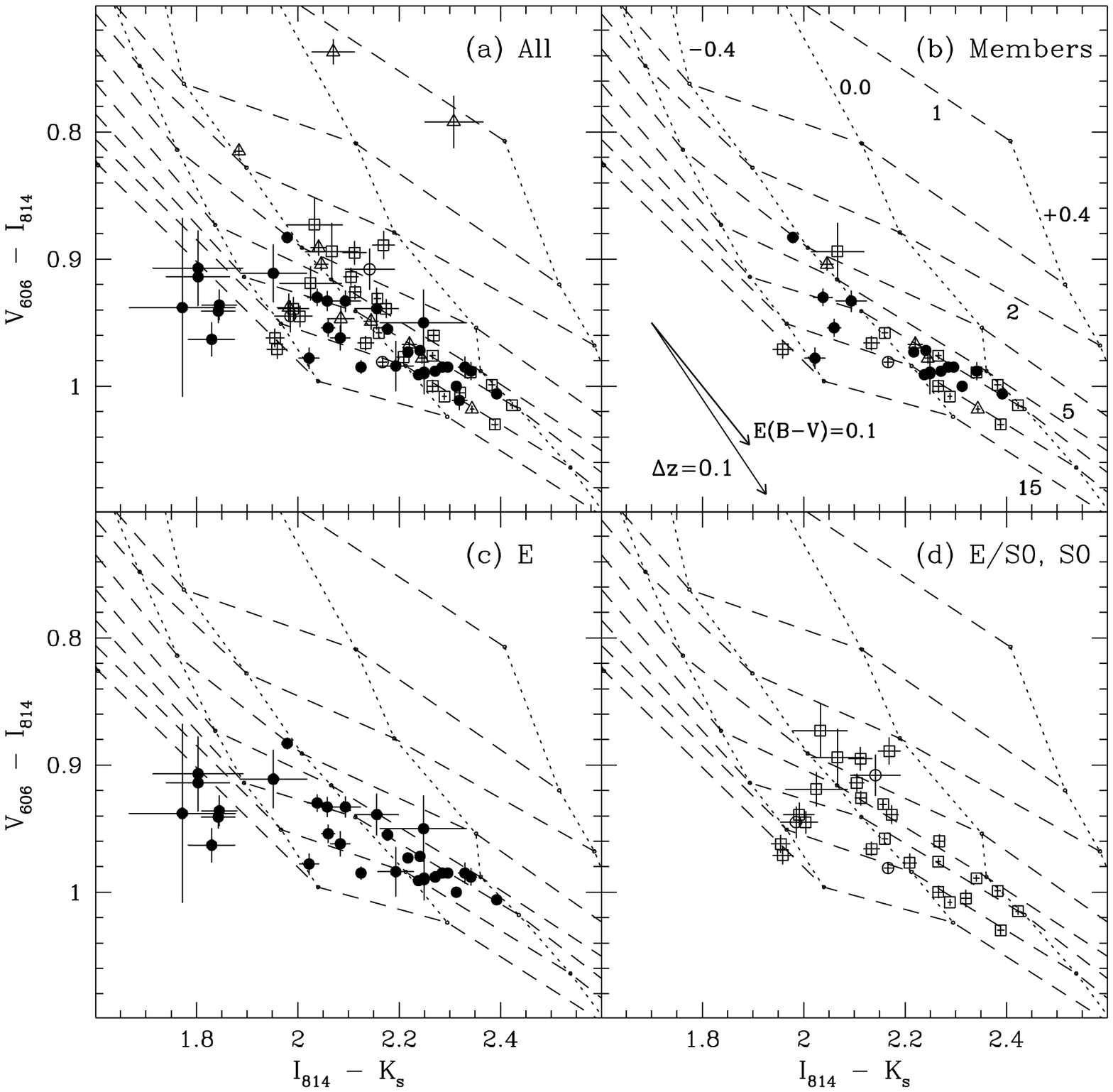,width=6.0in}}
\caption{A mosaic of colour--colour plots illustrating their
distribution in $(V_{606}-I_{814})$--$(I_{814}-K_{s})$ plane for the
$K_{s}<19$ sample.  The points are coded on the basis of their optical
morphology using the same symbols as Figure~3.  The four panels show:  a)
the whole sample, b) just the spectroscopically-confirmed members; c)
only the morphologically-typed ellipticals, d) only the morphologically
classified S0s and E/S0.  The relatively luminosities of the various
galaxies can be gauged from the size of the errorbars on the data points.}
\end{figure*}

\section{Results and Discussion}

We show the three colour--magnitude diagrams for our sample in Figure~2.
These illustrate the wide range in luminosity covered in our analysis,
nearly 6 magnitudes, with the elliptical and lenticular galaxies evenly
spread across this range, except for the brightest two
galaxies which are both ellipticals.  This figure also shows the relatively
homogeneous colours of the majority of the galaxies in our sample.
The combined elliptical and lenticular sample exhibits a scatter around
their mean relations of 0.11\,mag in $(B_{450}-I_{814})$, 0.17\,mag in
$(I_{814}-K_s)$ and a mere 0.03\,mag in $(V_{606}-I_{814})$.  However,
this scatter is still significantly larger than the mean errors in these
colours: $\delta(B_{450}-I_{814})=0.02$, $\delta(I_{814}-K_s)=0.03$
and $\delta(V_{606}-I_{814})=0.01$.  As we show next, the high precision of
our photometry of the early-type galaxies in A\,2218 means that we can
analyse the small scatter shown in their colours to uncover differences
in their stellar populations.

In Figure~3 we show the two colour--colour diagrams based on the
$(B_{450}-I_{814})$ and $(V_{606}-I_{814})$ comparisons with
$(I_{814}-K_s)$.  We overplot on this figure the grid of models for a
range of ages and metallicities (as described in \S2.5).  We indicate
in both panels the effects on the observed colours of increasing
reddening and redshift (for early-type SEDs).  The redshift vector
suggests that the fainter end of our sample could suffer field
contamination from foreground low-luminosity, but metal-rich,
early-type galaxies.  This combination of properties is sufficiently
rare and the foreground volume small enough that we do not expect
contamination to be a major problem in our analysis.

A comparison of the model grids and the observations in Figure~3
shows good agreement between the regions of the colour--colour plane
populated by the observations and those where galaxies are expected to
lie on the basis of the stellar population models.  However, comparing
the `orthogonality' of the age and metallicity tracks on the two
colour--colour planes in Figure~3 we can see that the $(V_{606}-I_{814})$
grid provides a better differentiation between age and metallicity
than  $(B_{450}-I_{814})$ as expected from the discussion in \S2.5.
Hence, for the remainder of our analysis we will concentrate on the
$(V_{606}-I_{814})$--$(I_{814}-K_s)$ colour--colour plane and compare
the detailed distribution of the galaxies on it to the model tracks.

The dominant feature in the $(V_{606}-I_{814})$ panel in Figure~3 (and
also shown by the $(B_{450}-I_{814})$ plot) is a tight locus of points
in the region of the grid corresponding to the oldest and most metal
rich stellar populations.  These galaxies are typically the brightest
early-type galaxies in the cluster (Fig.~2).  Their locus traces what
appears to be a metallicity sequence -- spanning a range of $\sim
0.5$\,dex in [M/H] -- at a single age ($\sim 7$\,Gyrs -- although the
uncertainty in the metallicities and the relative calibration of the
colours and grids means that the reader should view the ages as defined
on a relative rather than an absolute scale).  This can be more clearly
seen in Figure~4, which illustrates the
$(V_{606}-I_{814})$--$(I_{814}-K_s)$ distributions for the
morphologically classified elliptical and lenticular galaxies (we place
the three E/S0 galaxies into the latter sample).  The fainter galaxies
(identified by the larger photometric errors) extend this sequence to
lower [M/H], but also appear to show a wider spread in ages.  Indeed a
number of the faintest elliptical galaxies fall just outside the region
covered by the model predictions, although their photometric errors are
sufficiently large that we do not view this as a major concern.
Unfortunately none of these galaxies have spectroscopic information to
confirm their cluster membership.

To confirm the reliability of our analysis we compare our results for
the 14 galaxies in common with the spectroscopic sample of Ziegler et
al.\ (2000).   We note that these galaxies are typically the brighter
ones in our sample, with a median magnitude of $K_s=15.1$, the faintest
having $K_s=16.3$ ($0.25 L^\ast_K$).  We find that most of the galaxies
are consistent in both analyses with a single age ($\sim 7$--8\,Gyrs)
and a range of metallicity.  However, one galaxy stands out in our
sample -- \#449, the bluest, confirmed elliptical member in Figure~4b
-- the luminosity-weighted age we derive for the stellar population in
this galaxy is a mere 3\,Gyr (suggesting that the stellar population
which currently dominates the galaxy's luminosity was formed at a
redshift of only $z \ls 0.5$).  This galaxy also stands out in Ziegler
et al.'s analysis as by far the youngest galaxy in the joint sample,
with an estimated age of around $\sim 1.5$\,Gyrs. This comparison with
the results of the more traditional spectroscopic approach clearly
supports the reliability of the optical/near-infrared photometric
technique for analysing the stellar populations of early-type galaxies
in distant clusters.  Moreover, the photometric method requires only a
fraction of the time needed for  spectroscopic surveys and can probe to
fainter luminosities.

Turning back to Figure~4 we now compare the colours of the different
morphological subsamples.  The sequence defined by the redder and more
luminous galaxies appears in both the E and E/S0+S0 subsamples, as well
as in the spectroscopically-confirmed members.  There maybe a slight
tendency for the bright E/S0+S0 sample to show a wider age spread than
the E's ($\sigma_T({\rm S0})=2.2$\,Gyrs versus $\sigma_T({\rm
E})=0.8$\,Gyrs at $K_s<16.0$). The likelihood of this
difference in dispersions arising by chance  from the joint
sample is only slight, $\log_{10}(P)=-3.1$. However, this 
comparison is very sensitive to the exact magnitude limit used.

Nevertheless, there is an even more striking difference between the two
morphological subsamples at the lowest luminosities.  To emphasise the
variation in the properties of the lowest luminosity galaxies we show
in Figure~5 the colour--colour plot for the fainter ($K_s>17$, $\ls 0.1
L^\ast_K$) early-type galaxies.  Here, while the faint ellipticals
appear to extend the metallicity sequence defined by the brighter
ellipticals (and S0's), the fainter S0 galaxies show an almost
orthogonal colour--colour relationship to that exhibited by the faint
ellipticals and all the more luminous galaxies (qualitatively similar
behaviour is seen in the equivalent $(B_{450}-I_{814})$ plot).  This
reversal in the colour trends results from the appearance of a
population of faint S0 with relatively blue $(V_{606}-I_{814})$
colours, indicating luminosity-weighted mean ages of only 2--5\,Gyrs
for the stellar populations in these galaxies and metallicities of
[M/H]\,$\sim-0.2$.  It is unlikely that all of the stars in these
galaxies were formed in the most recent star formation event and hence
the luminosity-weighted ages are in fact upper limits on the epoch of
the most recent star formation.

To quantify the prevelance of this population we arbitrarily define
`young' as a luminosity-weighted age of $<5$\,Gyrs and  find that half
of the $K_s>17$ S0 galaxies are young (or 30\% of the whole S0
population, Fig.~6), in contrast none of the faint E's fall in this
category.  This is a high enough proportion of the S0 population (given
the duration of the `young' phase) to suggest that most of the faint
lenticular galaxies in the cluster must have passed through this
phase.  To quantify the extent of any possible luminosity bias in our
analysis we determine that the fading by the present-day of the stellar
populations of the young S0 galaxies  should not exceed 0.2\,mags in
$K_s$ compared to the more luminous, evolved galaxies in the cluster
(Smail et al.\ 1998; Kodama \& Bower 2000).  Taking this modest
luminosity bias into account we find that it would reduce the
proportion of `young' S0s in an unbiased sample by only 5\%.

%
%
\centerline{\psfig{file=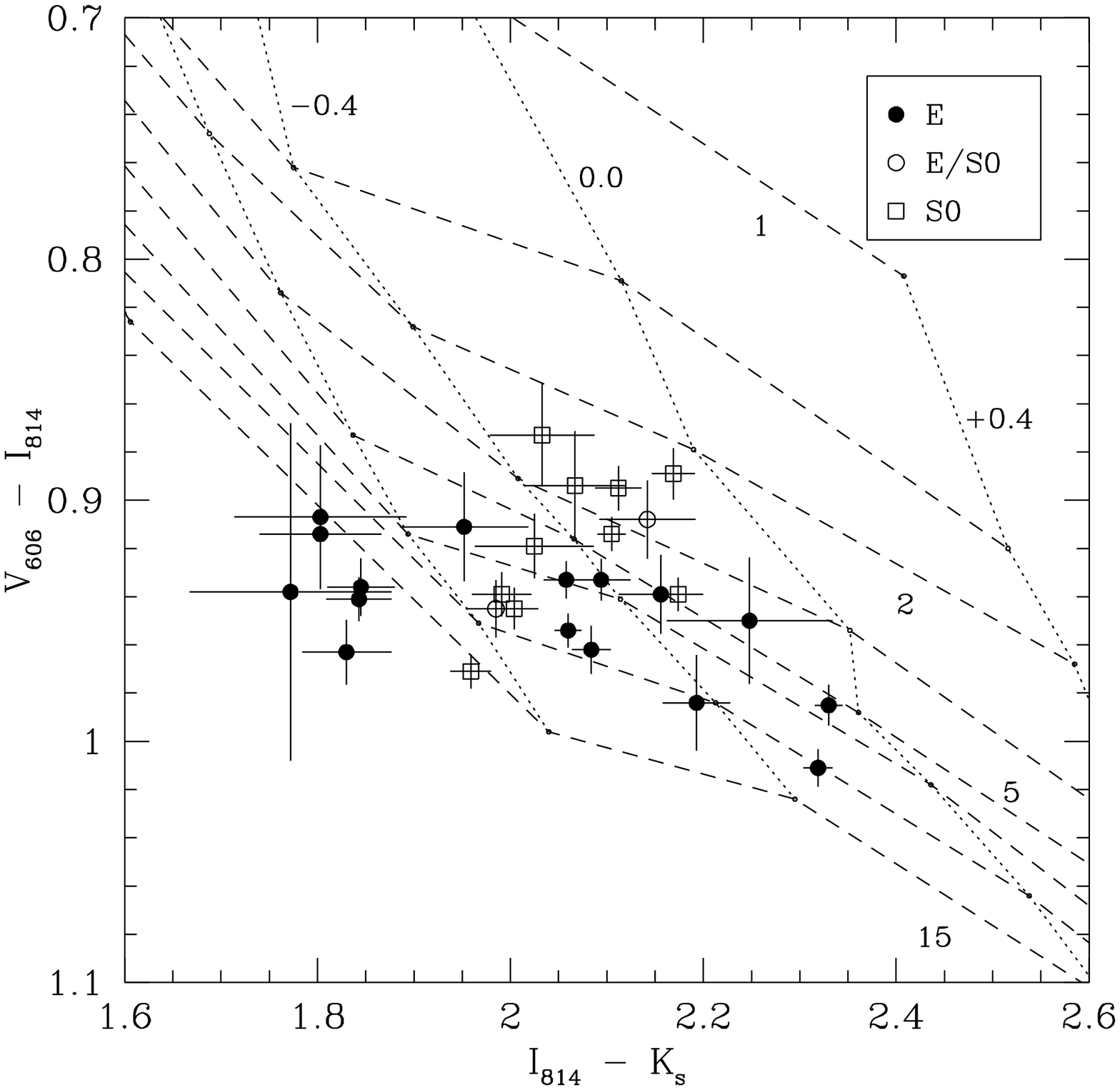,angle=0,width=3.0in}}
\noindent{{\bf Figure 5.} This plot compares the distribution of the
faintest early-type galaxies ($K_s>17$) in our sample  to the
predictions of the stellar population models.   The striking difference
in the distribution of faint E and S0 galaxies is obvious.  }

The behaviour exhibited by the faint lenticular population is very
similar to that seen in the `UV+' galaxies identified in a photometric
survey of X-ray luminous clusters at $z\sim 0.2$ by Smail et
al.\ (1998).   Their relatively shallow ground-based imaging lacked the
precision necessary for an analysis of the type undertaken here.
Nevertheless, they used the increased sensitivity to differences in the
stellar populations available from their $(U-B)$ colours to
identify a class of UV-bright galaxies which showed strong
4000\AA\ breaks (`UV+') indicative of evolved stellar populations.
They suggested that these UV+ galaxies represented the fading
progenitors of the cluster S0 population.  The UV+ class constitute
around 50\% of the faint cluster population ($<0.1L^\ast_V$) which
exhibit strong 4000\AA\ breaks.  Based on the {\it HST}\, imaging of a
handful of the UV+ galaxies in the core of A\,2390, Smail et
al.\ (1998) suggest that the majority of these galaxies have S0
morphologies.  The luminosities, colours and morphologies of the UV+
population are thus similar to the faint, blue S0 population identified
here.  Interestingly, Smail et al.\ (1998) found that the radial
profile of the UV+ population is indistinguishable from that of the
luminous, evolved galaxies in these clusters suggesting that these
galaxies are not recent additions to the cluster population.

We finally discuss the characteristics of the S0 population in the context
of the morphological evolution discussed by Dressler et al.\ (1997) and
Poggianti et al.\ (1999).  The evolution in the number of S0 galaxies in
clusters at $z>0.3$--0.6 observed by Dressler et al.\ (1997) suggests that
roughly two-thirds of the S0 population in a $z=0.17$ cluster would have
transformed their morphologies in the last 3\,Gyrs (since $z\sim 0.5$),
with half of these occuring in the last 1--2\,Gyrs.  Adopting a 1--2\,Gyr
delay between the termination of star formation and the morphological
transformation occuring, as is needed to explain the lack of blue
S0 galaxies in the distant clusters and the large number of passive
galaxies with late-type morphologies (Poggianti et al.\ 1999; Kodama \&
Smail 2000), then we would expect that  60--70\% of the S0 galaxies in
A\,2218  were forming stars 4--5\,Gyrs prior to $z=0.17$ and 30\% of
them within the last 2--3\,Gyrs.   To estimate the luminosity-weighted
ages for the composite stellar populations in these galaxies at a later
time we need to assume a model for their previous star formation. If
we assume that these progenitors were forming stars at a constant rate
(or slightly declining) since their formation at high redshift, as normal
early and mid-type spiral galaxies are believed to do, then for the 30\%
of the S0 population which is predicted to have had terminated its star
formation in the  2--3\,Gyrs before $z=0.17$ (equivalent to $z\sim 0.5$
in the cosmology we adopt) they would have luminosity-weighted ages of
$\ls 5$\,Gyrs when seen at $z=0.17$.  This model is thus consistent
with the 30\% fraction of the S0 population which we observe to have
luminosity-weighted ages of $\ls 5$\,Gyrs.  Indeed, the smooth growth
of the S0 fraction in the clusters reported by Dressler et al.\ (1997)
is also mirrored by the nearly flat distribution of estimated ages for
this population seen in Figure~6.  Therefore, our observations support a
scenario where the fraction of S0s in the clusters is smoothly increasing
as a result of a slow morphological and spectroscopic transformation of
star forming spiral galaxies into passive, S0 cluster members.

%
%
\centerline{\psfig{file=fig6.ps,angle=270,width=3.0in}}
\noindent{{\bf Figure 6.} The cumulative distributions of the estimated
luminosity-weighted relative ages of the full elliptical and lenticular
samples.   Note the relatively flat distribution of derived ages for the
S0+E/S0 sample compared to the more uniform and older ages seen in the
ellipticals, where the mean stellar populations in 80\% of the galaxies
are older than 7\,Gyrs. We stress that these estimated ages should be
viewed as relative rather than absolute values.   }

\section{Conclusions}

\noindent{$\bullet$}  We present precise optical and near-infrared
photometry of a sample of 81 galaxies in the core regions of the rich
cluster A\,2218 at $z=0.17$ (a lookback time of 3\,Gyrs).

\noindent{$\bullet$} We compare the optical and optical-infrared
colours of the morphologically-classified early-type galaxies in our
sample with the predictions of stellar population models spanning a
range in metallicity and age.  The models  show good agreement in the
regions of the colour--colour plane which are populated by the
observational data.

\noindent{$\bullet$} By comparing our results with those from a recent
spectroscopic survey of the stellar populations in galaxies in A\,2218,
we confirm the reliability of the photometric analysis technique.  We
show that both techniques identify galaxies with relatively young
stellar populations.

\noindent{$\bullet$} The colours of the more luminous early-type galaxies
($\gs 0.5L^\ast_K$) in the cluster are well described by a sequence
of varying metallicity for the stellar population at a constant age.
This is in agreement with the results of the analysis of the spectral
line strengths in similar luminosity galaxies in A\,2218 (Ziegler et
al.\ 2000) and results from the spectroscopic and photometric analysis
of galaxies in higher redshift clusters (Kodama \& Arimoto 1997; Jones,
Smail \& Couch 2000).

\noindent{$\bullet$} In contrast the faintest early-type galaxies ($\ls
0.1L^\ast_K$) show a large spread in their optical and optical-infrared
colours.  Around 30\% of the S0 galaxies in A\,2218 (and half of the
fainter examples) exhibit mean luminosity-weighted ages for their stellar
populations of $\ls 5$\,Gyrs.  This suggests that these galaxies were
actively forming stars at $z\ls 0.5$.  The proportion of these `young'
S0 galaxies is consistent with the rate of evolution of the S0 fraction
in distant clusters reported by Dressler et al.\ (1997).  Further support
for the gradual growth of the S0 population comes from the relatively
flat age distribution we derive for these galaxies.  In contrast the
vast majority of the morphologically-classified elliptical galaxies
across all luminosities show evolved stellar populations.

\noindent{$\bullet$} Most of the galaxies we see which show signs of
recent activity lie below the magnitude limits reached by spectroscopic
studies of galaxies in distant clusters.  This highlights the urgent
need to exploit the increased sensitivity available with 8-m telescopes
to measure detailed line indices in low-luminosity members of intermediate
and moderate redshift clusters.

\noindent{$\bullet$}  We have demonstrated the power of photometric
analysis of the optical--infrared colours of early-type cluster galaxies
to uncover information about their star formation histories.  This type
of precision photometric analysis should be extended to more distant
clusters (in which it is predicted that a higher fraction of the S0
galaxies should show signs of past star formation activity) and to wider
fields at intermediate redshifts to search for the signatures of the
physical processes responsible for the spectroscopic and morphological
transformations at the heart of the formation of S0 galaxies.

\section*{Acknowledgements}

We thank Richard Bower, Warrick Couch, Roger Davies, Bianca Poggianti,
Ray Sharples and Alex Vazdekis for useful conversations and help.  IRS
acknowledges support from the Royal Society.  HK and GPS acknowledge
support from PPARC.  TK acknowledges support through a Research
Fellowship for Young Scientists from the Japan Society for the
Promotion of Science.

%
%
\newpage

\setcounter{table}{0}
\begin{table*}
\begin{center}
\caption{\hfil Galaxy Catalogue \hfil}
\begin{tabular}{lccccccclcl}
{ID} & {ID$_{\rm spec}^a$} & {$\alpha$,$\delta$~(J2000)} & {$K_s^{tot}$} & {$(B_{450}-I_{814})$} & {$(V_{606}-I_{814})$} & {$(I_{814}-K_{s})$} &  {$r_{hl}$} & {Morph} & {$z$} & {Note$^b$} \cr
     &                   & 16$^{\rm h}$, +66$^\circ$ &             &                     &                     &
                    & ($''$)      &       &     &           \cr
\noalign{\medskip}
 103  & ...  &  35\,48.48, 12\, 1.7  & $17.72\pm 0.05$ & $2.08\pm 0.03$ & $0.79\pm 0.02$ & $2.31\pm 0.06$ & 1.05  &     Sdm  &   ...  & \cr	
 129  & ...  &  35\,46.60, 12\,27.0  & $17.61\pm 0.03$ & $2.33\pm 0.02$ & $0.94\pm 0.01$ & $1.98\pm 0.02$ & 0.41  &    S0/a  &   ...  & \cr	
 131  &  L404  &  35\,48.80, 12\, 9.7  & $16.93\pm 0.02$ & $2.31\pm 0.01$ & $0.93\pm 0.01$ & $2.04\pm 0.02$ & 0.70  &       E  &  0.1708  & \cr	
 137  &  Z1516  &  35\,46.84, 12\,22.9  & $15.52\pm 0.01$ & $2.44\pm 0.01$ & $0.98\pm 0.00$ & $2.24\pm 0.01$ & 0.77  &   SB0/a  &  0.1638  & \cr	
 145  & ...  &  35\,50.63, 12\, 1.9  & $16.82\pm 0.01$ & $2.41\pm 0.01$ & $0.95\pm 0.01$ & $2.15\pm 0.01$ & 0.53  &     Sab  &   ...  & \cr	
 151  & ...  &  35\,50.23, 12\, 5.6  & $18.24\pm 0.04$ & $2.30\pm 0.02$ & $0.94\pm 0.01$ & $1.99\pm 0.03$ & 0.37  &    E/S0  &   ...  & \cr	
 153  & ...  &  35\,52.49, 11\,50.6  & $17.21\pm 0.02$ & $2.55\pm 0.02$ & $0.98\pm 0.01$ & $2.33\pm 0.01$ & 0.28  &       E  &   ...  & \cr	
 154  &  B143  &  35\,45.99, 12\,32.9  & $16.54\pm 0.01$ & $2.51\pm 0.01$ & $0.99\pm 0.01$ & $2.34\pm 0.01$ & 0.43  &       E  &  0.1659  & \cr	
 159  & ...  &  35\,45.02, 12\,44.7  & $17.72\pm 0.03$ & $1.95\pm 0.02$ & $0.89\pm 0.01$ & $2.17\pm 0.02$ & 0.36  &      S0  &   ...  & \cr	
 191  & ...  &  35\,51.32, 12\, 7.6  & $17.30\pm 0.03$ & $2.27\pm 0.02$ & $0.90\pm 0.01$ & $2.11\pm 0.02$ & 0.62  &      S0  &   ...  & \cr	
 195  & ...  &  35\,44.97, 12\,53.7  & $17.80\pm 0.05$ & $2.27\pm 0.02$ & $0.94\pm 0.01$ & $1.84\pm 0.04$ & 0.65  &       E  &   ...  & \cr	
 205  &  B031  &  35\,50.78, 12\,20.5  & $15.21\pm 0.01$ & $2.60\pm 0.01$ & $1.03\pm 0.00$ & $2.39\pm 0.01$ & 0.75  &      S0  &  0.1835  & \cr	
 220  & ...  &  35\,40.87, 12\,36.7  & $17.08\pm 0.02$ & $2.35\pm 0.01$ & $0.94\pm 0.01$ & $2.17\pm 0.01$ & 0.52  &      S0  &   ...  & \cr	
 230  & ...  &  35\,47.60, 12\,42.8  & $15.15\pm 0.01$ & $2.54\pm 0.01$ & $1.00\pm 0.01$ & $2.32\pm 0.01$ & 0.81  &      S0  &   ...  & \cr	
 232  & ...  &  35\,46.33, 12\,52.0  & $15.39\pm 0.02$ & $2.38\pm 0.01$ & $0.96\pm 0.00$ & $2.18\pm 0.01$ & 0.80  &       E  &   ...  & \cr	
 267  & ...  &  35\,47.41, 13\, 5.3  & $17.96\pm 0.06$ & $2.55\pm 0.03$ & $0.98\pm 0.02$ & $2.19\pm 0.03$ & 0.23  &       E  &   ...  & \cr	
 279  & ...  &  35\,50.25, 12\,19.6  & $16.47\pm 0.01$ & $2.25\pm 0.01$ & $0.93\pm 0.01$ & $2.11\pm 0.01$ & 0.69  &      S0  &   ...  & \cr	
 280  &  B024  &  35\,50.00, 12\,23.8  & $14.37\pm 0.01$ & $2.55\pm 0.01$ & $1.02\pm 0.00$ & $2.34\pm 0.00$ & 1.03  &   SB0/a  &  0.1776  &  Z1552 \cr	
 285  &  L357  &  35\,51.11, 12\,34.1  & $16.77\pm 0.04$ & $2.55\pm 0.02$ & $0.99\pm 0.02$ & $2.25\pm 0.02$ & 0.32  &       E  &  0.1730  & \cr	
 292  &  L259  &  35\,56.33, 11\,51.1  & $15.11\pm 0.01$ & $2.46\pm 0.01$ & $0.98\pm 0.00$ & $2.27\pm 0.01$ & 0.91  &      S0  &  0.1646  & \cr	
 298  &  B020  &  35\,49.47, 12\,36.3  & $14.96\pm 0.01$ & $2.42\pm 0.01$ & $0.96\pm 0.00$ & $2.16\pm 0.01$ & 0.88  &      S0  &  0.1751  &  Z1580 \cr	
 301  &  K1  &  35\,49.31, 12\,44.7  & $13.19\pm 0.00$ & $2.56\pm 0.00$ & $0.97\pm 0.00$ & $2.24\pm 0.00$ & 3.50  &       E  &  0.1720  &  cD \cr	
 303  &  B117  &  35\,55.92 12\, 3.4  & $16.98\pm 0.02$ & $2.45\pm 0.02$ & $0.98\pm 0.01$ & $2.02\pm 0.02$ & 0.69  &       E  &  0.1775  & \cr	
 307  &  B003  &  35\,56.81, 11\,55.5  & $13.77\pm 0.00$ & $2.54\pm 0.00$ & $0.99\pm 0.00$ & $2.24\pm 0.00$ & 2.84  &       E  &  0.1768  &  Z1437 \cr	
 315  &  B018  &  35\,51.86, 12\,34.3  & $14.55\pm 0.00$ & $2.52\pm 0.01$ & $1.01\pm 0.00$ & $2.39\pm 0.00$ & 0.71  &       E  &  0.1637  &  Z1662 \cr	
 323  & ...  &  35\,59.05, 11\,47.5  & $16.82\pm 0.01$ & $2.28\pm 0.02$ & $1.03\pm 0.01$ & $2.64\pm 0.02$ & 0.58  &      Sc  &   ...  & \cr	
 324  & ...  &  35\,54.90, 12\,18.1  & $16.83\pm 0.02$ & $2.23\pm 0.01$ & $0.89\pm 0.01$ & $2.04\pm 0.02$ & 0.71  &     Sab  &   ...  & \cr	
 326  &  B070  &  35\,53.38, 12\,38.6  & $16.03\pm 0.01$ & $2.32\pm 0.01$ & $0.90\pm 0.01$ & $2.05\pm 0.01$ & 1.16  &     Sab  &  0.1791  & \cr	
 333  &  B048  &  35\,49.05, 13\, 0.8  & $15.64\pm 0.01$ & $1.67\pm 0.01$ & $0.82\pm 0.00$ & $1.88\pm 0.01$ & 0.92  &      Sc  &  0.1032  & \cr	
 337  &  B039  &  35\,47.25, 13\,16.1  & $15.20\pm 0.01$ & $2.56\pm 0.01$ & $0.99\pm 0.00$ & $2.28\pm 0.01$ & 0.78  &       E  &  0.1800  &  Z2604 \cr	
 350  & ...  &  35\,43.50, 12\,21.0  & $18.85\pm 0.08$ & $2.28\pm 0.03$ & $0.91\pm 0.02$ & $2.14\pm 0.05$ & 0.40  &    E/S0  &   ...  & \cr	
 359  & ...  &  35\,51.23, 12\,55.0  & $17.23\pm 0.03$ & $2.41\pm 0.02$ & $0.96\pm 0.01$ & $2.08\pm 0.02$ & 0.53  &       E  &   ...  & \cr	
 368  & ...  &  35\,59.12, 12\, 1.3  & $16.86\pm 0.03$ & $2.37\pm 0.02$ & $1.21\pm 0.01$ & $2.54\pm 0.02$ & 0.70  &      Sc  &   ...  & \cr	
 376  &  B064  &  35\,57.43, 12\,15.7  & $15.54\pm 0.01$ & $2.55\pm 0.01$ & $1.00\pm 0.00$ & $2.38\pm 0.01$ & 0.51  &      S0  &  0.1820  &  Z1454 \cr	
 380  & ...  &  35\,51.06, 13\, 3.3  & $18.45\pm 0.10$ & $2.47\pm 0.04$ & $0.87\pm 0.02$ & $2.03\pm 0.05$ & 0.62  &      S0  &   ...  & \cr	
 387  & ...  &  35\,56.49, 12\,27.1  & $18.79\pm 0.05$ & $2.39\pm 0.03$ & $0.96\pm 0.01$ & $1.83\pm 0.05$ & 0.27  &       E  &   ...  & \cr	
 398  & ...  &  36\, 1.26, 11\,55.8  & $17.93\pm 0.06$ & $2.35\pm 0.03$ & $0.94\pm 0.02$ & $2.16\pm 0.04$ & 0.56  &       E  &   ...  & \cr	
 401  &  B030  &  35\,59.42, 12\, 6.6  & $14.85\pm 0.01$ & $2.51\pm 0.01$ & $0.99\pm 0.00$ & $2.27\pm 0.01$ & 1.29  &       E  &  0.1798  &  Z1466 \cr
 420  & ...  &  35\,53.49, 12\,57.2  & $16.52\pm 0.01$ & $3.10\pm 0.03$ & $1.21\pm 0.01$ & $2.89\pm 0.01$ & 0.82  &     Scd  &   ...  & \cr	
 424  & ...  &  36\, 2.10, 11\,58.4  & $18.93\pm 0.11$ & $2.33\pm 0.03$ & $0.91\pm 0.02$ & $1.80\pm 0.06$ & 0.33  &       E  &   ...  & \cr	
 428  &  B028  &  36\, 2.33, 11\,52.8  & $14.60\pm 0.00$ & $2.60\pm 0.01$ & $1.01\pm 0.00$ & $2.42\pm 0.00$ & 0.78  &      S0  &  0.1830  &  Z1343 \cr
 439  & ...  &  36\, 0.57, 12\, 7.1  & $16.57\pm 0.02$ & $2.48\pm 0.01$ & $0.98\pm 0.01$ & $2.21\pm 0.01$ & 0.60  &      S0  &   ...  & \cr	
 449  &  B087  &  35\,56.72, 12\,41.3  & $16.27\pm 0.01$ & $2.17\pm 0.01$ & $0.88\pm 0.00$ & $1.98\pm 0.01$ & 0.73  &       E  &  0.1717  &  Z1605 \cr
 457  &  B121  &  35\,54.67, 13\, 2.4  & $18.97\pm 0.10$ & $2.42\pm 0.05$ & $0.89\pm 0.02$ & $2.07\pm 0.05$ & 0.47  &    S0/a  &  0.1717  & \cr	
 470  & ...  &  35\,45.46, 12\, 8.5  & $17.16\pm 0.02$ & $2.28\pm 0.01$ & $0.91\pm 0.01$ & $2.11\pm 0.01$ & 0.43  &      S0  &   ...  & \cr	
 503  &  B047  &  35\,59.38, 12\,53.6  & $15.38\pm 0.01$ & $2.46\pm 0.01$ & $0.97\pm 0.00$ & $2.22\pm 0.01$ & 0.79  &       E  &  0.1747  &  Z1711 \cr
 537  &  Z2660  &  35\,58.39, 13\,21.5  & $16.05\pm 0.01$ & $2.52\pm 0.01$ & $0.98\pm 0.00$ & $2.17\pm 0.01$ & 0.73  &    E/S0  &  0.1773  & \cr
 545  & ...  &  36\, 0.96, 13\,10.2  & $18.76\pm 0.08$ & $2.14\pm 0.03$ & $0.92\pm 0.01$ & $2.02\pm 0.06$ & 0.83  &      S0  &   ...  & \cr	
 563  &  B152  &  35\,59.27, 13\,33.8  & $17.16\pm 0.02$ & $2.39\pm 0.01$ & $0.95\pm 0.01$ & $2.06\pm 0.01$ & 0.40  &       E  &  0.1787  & \cr	
 571  & ...  &  36\, 0.00, 13\,33.5  & $18.35\pm 0.06$ & $3.12\pm 0.09$ & $1.08\pm 0.03$ & $2.96\pm 0.06$ & 0.87  &     Scd  &   ...  & \cr	
 599  &  B159  &  36\, 3.81, 13\,14.3  & $17.32\pm 0.02$ & $2.44\pm 0.01$ & $0.97\pm 0.01$ & $1.96\pm 0.02$ & 0.53  &      S0  &  0.1731  & \cr	
 612  &  L143  &  36\, 2.64, 13\,32.3  & $16.55\pm 0.02$ & $2.44\pm 0.01$ & $0.97\pm 0.01$ & $2.13\pm 0.02$ & 0.82  &    S0/a  &  0.1811  & \cr	
 633  &  B017  &  36\, 4.25, 13\,25.3  & $14.88\pm 0.00$ & $2.57\pm 0.01$ & $1.01\pm 0.00$ & $2.29\pm 0.00$ & 0.70  &    S0  &  0.1738  &  Z2702 \cr
 634  & ...  &  36\, 4.40, 13\,28.7  & $16.33\pm 0.01$ & $2.48\pm 0.01$ & $0.98\pm 0.01$ & $2.13\pm 0.01$ & 0.64  &       E  &   ...  & \cr	
 638  & ...  &  36\, 3.81, 13\,33.9  & $15.75\pm 0.01$ & $2.31\pm 0.01$ & $0.93\pm 0.00$ & $2.16\pm 0.01$ & 0.71  &      S0  &   ...  & \cr	
 646  & ...  &  36\, 2.52, 13\,50.8  & $18.32\pm 0.06$ & $2.64\pm 0.05$ & $1.28\pm 0.02$ & $2.95\pm 0.05$ & 0.56  & Sdm  &   ...  & \cr	
 647  & ...  &  36\, 4.75, 13\,33.9  & $17.97\pm 0.03$ & $2.36\pm 0.02$ & $0.94\pm 0.01$ & $1.84\pm 0.03$ & 0.31  &       E  &   ...  & \cr	
 660  & ...  &  35\,43.40, 12\,28.4  & $17.94\pm 0.04$ & $2.36\pm 0.02$ & $0.95\pm 0.01$ & $2.09\pm 0.03$ & 0.50  &      Sb  &   ...  & \cr	
\noalign{\smallskip}
\end{tabular}
\end{center}
$a$) Z: Ziegler et al.\ (2000), B: Butcher et al.\ (1983), L: Le~Borgne et al.\ (1992), K: Kristian et al.\ (1978) -- 
$b$) Ziegler et al.\ (2000). 
\end{table*}

\newpage
%
%
\setcounter{table}{0}
\begin{table*}
\begin{center}
\caption{\hfil Galaxy Catalogue (cont) \hfil}
\begin{tabular}{lccccccclcl}
{ID} & {ID$_{\rm spec}^a$} & {$\alpha$,$\delta$~(J2000)} & {$K_s^{tot}$} & {$(B_{450}-I_{814})$} & {$(V_{606}-I_{814})$} & {$(I_{814}-K_{s})$} &  {$r_{hl}$} & {Morph} & {$z$} &  {Note$^b$} \cr
     &                   & 16$^{\rm h}$, +66$^\circ$ &             &                     &                     &
                    & ($''$)      &       &     &           \cr
\noalign{\medskip}
1018  & ...  &  36\, 0.43, 12\,20.8  & $17.08\pm 0.03$ & $2.34\pm 0.02$ & $0.93\pm 0.01$ & $2.06\pm 0.02$ & 0.83  &       E  &   ...  & \cr	
1047  &  L187  &  36\, 0.34, 12\,44.2  & $15.77\pm 0.01$ & $2.51\pm 0.01$ & $1.00\pm 0.00$ & $2.31\pm 0.01$ & 0.46  &       E  &  0.1750  & \cr	
1057  &  B019  &  36\, 2.24, 12\,34.5  & $14.57\pm 0.00$ & $2.49\pm 0.01$ & $0.99\pm 0.00$ & $2.25\pm 0.00$ & 1.27  &       E  &  0.1753  & \cr	
1060  & ...  &  36\, 3.33, 12\,32.6  & $17.36\pm 0.03$ & $2.35\pm 0.02$ & $0.94\pm 0.01$ & $2.00\pm 0.02$ & 0.41  &      S0  &   ...  & \cr	
1066  & ...  &  36\, 3.29, 12\,36.7  & $17.51\pm 0.03$ & $2.36\pm 0.02$ & $0.94\pm 0.01$ & $1.99\pm 0.03$ & 0.71  &      S0  &   ...  & \cr	
2005  & ...  &  35\,47.47, 12\,24.2  & $18.53\pm 0.09$ & $2.27\pm 0.04$ & $0.91\pm 0.03$ & $1.80\pm 0.09$ & 0.44  &       E  &   ...  & \cr	
2010  & ...  &  35\,49.18, 12\,31.9  & $18.27\pm 0.13$ & $2.28\pm 0.07$ & $0.94\pm 0.07$ & $1.77\pm 0.10$ & 0.25  &       E  &   ...  & \cr	
2011  &  L389  &  35\,49.32, 12\,21.5  & $17.04\pm 0.04$ & $2.26\pm 0.01$ & $0.97\pm 0.01$ & $2.22\pm 0.01$ & 1.18  &      Sa  &  0.1800  & \cr	
4001  & ...  &  36\, 3.40, 12\, 3.2  & $18.93\pm 0.08$ & $1.20\pm 0.02$ & $0.53\pm 0.01$ & $1.42\pm 0.09$ & 0.53  &      Sd  &   ...  & \cr	
4002  & ...  &  36\, 4.22, 12\,12.0  & $15.99\pm 0.01$ & $2.42\pm 0.01$ & $0.96\pm 0.01$ & $2.27\pm 0.01$ & 0.92  &      S0  &   ...  & \cr	
4003  &  B185  &  36\, 3.54, 12\,26.5  & $17.46\pm 0.03$ & $2.32\pm 0.02$ & $0.93\pm 0.01$ & $2.09\pm 0.03$ & 0.68  &       E  &  0.1821  & \cr	
4004  & ...  &  36\, 6.45, 12\,29.3  & $17.78\pm 0.03$ & $>5.0 (2\sigma)$ & $3.00\pm 0.18$ & $4.30\pm 0.04$ & 0.29  &       ?  &   ...  & \cr	
4005  & ...  &  36\, 6.56, 12\,33.1  & $18.37\pm 0.04$ & $>4.1 (2\sigma)$ & $1.97\pm 0.17$ & $4.39\pm 0.08$ & 0.36  &      Sc  &   ...  & \cr	
4006  & ...  &  36\, 5.97, 12\,41.3  & $18.88\pm 0.08$ & $2.40\pm 0.05$ & $0.95\pm 0.03$ & $2.25\pm 0.09$ & 0.26  &       E  &   ...  & \cr	
4007  &  B036  &  36\, 6.41, 12\,47.8  & $15.17\pm 0.01$ & $2.52\pm 0.01$ & $0.99\pm 0.00$ & $2.34\pm 0.01$ & 0.79  &     SB0  &  0.1681  & \cr	
4008  & ...  &  36\, 5.64, 12\,49.4  & $17.88\pm 0.04$ & $1.54\pm 0.02$ & $0.74\pm 0.01$ & $2.07\pm 0.04$ & 0.73  &      Sb  &   ...  & \cr	
4009  & ...  &  36\, 7.43, 13\, 0.3  & $16.36\pm 0.01$ & $2.86\pm 0.02$ & $1.12\pm 0.01$ & $2.63\pm 0.01$ & 0.73  &      Sa  &   ...  & \cr	
4010  &  B175  &  36\, 4.14, 13\, 2.2  & $17.07\pm 0.02$ & $2.43\pm 0.02$ & $1.01\pm 0.01$ & $2.32\pm 0.02$ & 0.43  &       E  &  0.2913  & \cr	
4011  & ...  &  35\,43.63, 13\, 9.6  & $18.79\pm 0.08$ & $2.48\pm 0.04$ & $0.91\pm 0.02$ & $1.95\pm 0.07$ & 0.50  &       E  &   ...  & \cr	
4012  & ...  &  36\, 6.71, 13\,21.7  & $18.85\pm 0.06$ & $3.73\pm 0.20$ & $1.81\pm 0.05$ & $3.69\pm 0.06$ & 0.19  &       ?  &   ...  & \cr	
4013  &  B015  &  35\,44.12, 13\,20.4  & $14.52\pm 0.00$ & $2.57\pm 0.01$ & $0.98\pm 0.00$ & $2.30\pm 0.00$ & 0.96  &       E  &  0.1778  & \cr	
4014  &  Z2738  &  36\, 4.85, 13\,42.5  & $16.22\pm 0.01$ & $2.55\pm 0.01$ & $1.00\pm 0.01$ & $2.27\pm 0.01$ & 0.46  &      S0  &  0.1742  & \cr
4015  & ...  &  36\, 9.28, 13\,47.6  & $18.59\pm 0.04$ & $2.57\pm 0.04$ & $1.11\pm 0.02$ & $2.58\pm 0.05$ & 0.28  &      Sc  &   ...  & \cr	
\noalign{\smallskip}
\end{tabular}
\end{center}
$a$) Z: Ziegler et al.\ (2000), B: Butcher et al.\ (1983), L: Le~Borgne et al.\ (1992), K: Kristian et al.\ (1978) -- 
$b$) Ziegler et al.\ (2000). 
\end{table*}

\end{document}